\documentclass[pra,twocolumn,preprintnumbers,amsmath,amssymb,showpacs,superscriptaddress]
{revtex4}

\usepackage{graphicx}
\usepackage{amsfonts}
\usepackage{amssymb,amsmath}
\usepackage{dcolumn}
\usepackage{bm}
\usepackage{color}
\newcommand{\bra}[1]{\langle#1|}

\newcommand{\ket}[1]{|#1\rangle}
\providecommand{\openone}{\leavevmode\hbox{\small1\kern-3.8pt\normalsize1}}

\begin{document}

\title{Unified view of correlations using the square norm distance}

\author{Bruno Bellomo}
\affiliation{Dipartimento di Fisica, Universit\`a di Palermo, via Archirafi 36,
90123 Palermo, Italy}
\author{Gian Luca Giorgi}
\affiliation{IFISC (UIB-CSIC), Instituto de F\'{i}sica Interdisciplinar y-axes
Sistemas Complejos,
UIB Campus, E-07122 Palma de Mallorca, Spain}
\author{Fernando Galve}
\affiliation{IFISC (UIB-CSIC), Instituto de F\'{i}sica Interdisciplinar y
Sistemas Complejos,
UIB Campus, E-07122 Palma de Mallorca, Spain}
\author{Rosario Lo Franco}
\affiliation{Dipartimento di Fisica, Universit\`a di Palermo, via Archirafi 36,
90123 Palermo, Italy}
\affiliation{CSFNSM and Dipartimento di Fisica e Astronomia, Universit\`a di
Catania, Viale A. Doria 6, 95125 Catania, Italy}
\author{Giuseppe Compagno}
\affiliation{Dipartimento di Fisica, Universit\`a di Palermo, via Archirafi 36,
90123 Palermo, Italy}
\author{Roberta Zambrini}
\affiliation{IFISC (UIB-CSIC), Instituto de F\'{i}sica Interdisciplinar y
Sistemas Complejos,
UIB Campus, E-07122 Palma de Mallorca, Spain}

\begin{abstract}
The distance between a quantum state and its closest state not having a certain
property has been used to quantify the amount of correlations corresponding to
that property. This approach allows a unified view of the
various kinds of correlations present in a quantum system.
In particular, using relative entropy as a distance measure, total correlations
can be meaningfully separated in a quantum and a classical part thanks to an
additive relation involving only distances between states. Here, we investigate
a unified view of correlations using as distance measure the
square norm, already used to define the so-called geometric quantum discord. We
thus consider geometric quantifiers also for total and classical correlations
finding, for a quite general class of bipartite states, their explicit
expressions. We analyze the relationship among geometric total, quantum and
classical correlations and we find that they do not satisfy anymore a closed
additivity relation.
\end{abstract}

\pacs{03.67.Mn, 03.65.Ud, 03.65.Yz}

\maketitle

\section{Introduction}

Quantum systems have properties characterized by various kinds of
correlations some of which distinguish them from classical systems~\cite{epr,bell}. These properties may be an essential resource for
quantum computation and quantum information~\cite{{nielsen-chuang}}. This makes
important to distinguish among those correlations that are peculiar of quantum
systems with respect to the ones present also in classical systems. A nonlocal
property of quantum systems, entanglement, allows to achieve  exponential
speed-up in pure-state computation if it grows with the size of the
system~\cite{linden}. On the other hand, in the case of mixed-state computation,
in certain computational tasks quantum speed-up can be achieved using separable
(unentangled) states, like in the so-called deterministic quantum computation
with one qubit (DQC1) protocol~\cite{DQC1}. This speed-up has been
linked~\cite{datta} to the presence of quantum discord~\cite{zurek,henderson},
considered as a quantifier of the quantum part of correlations present in a
bipartite system and defined as the difference between two quantum analogues of
the classical mutual information~\cite{serra2011,laflamme2011}.

In the general case of a multipartite system, the various kinds of correlations
present in a quantum state have been linked to the distance between the system
state itself and its closest states without the desired property, allowing to
look at them in a unified view~\cite{Vedral1997PRL,Modi2010PRL}. Relative
entropy, although not symmetrical under the exchange of the entries, has
been used as distance measure between states. In this approach, the
decomposition of the total correlations, $T$,  in a classical, $C$, and a
quantum part, $D$, appears meaningful because $T$ equals the sum of $D$ and $C$
up to a quantity  $L$, which in turn is a relative entropy-based (REB) distance
between two of the relevant closest states. The quantity $L$ results to be, in the bipartite
case, equal to the difference between REB quantum discord and its original
definition~\cite{zurek,henderson}. Using relative entropy as a distance measure,
correlation quantifiers therefore satisfy a closed additivity relation among
them~\cite{Modi2010PRL}.

The properties of quantum discord have been widely investigated in the last
years~\cite{celeri2011}. It has been shown that it is present in almost all quantum
states~\cite{acin2010} and the relation between discord and entanglement has
been discussed \cite{cornelio,streltsov,piani}. In contrast with entanglement,
discord can be generated using local noise~\cite{noisediscord} and it is not
monogamous~\cite{monogamy}. Differently from what happens for entanglement,
quantum discord does not present sudden death~\cite{werlang} during its
evolution but can still present revivals
even in absence of system-environment back-action~\cite{revival}.
Generalizations of discord to the multipartite case have been also reported
following different approaches~\cite{multi}.

On the other hand, both the original and the REB discord require involved
minimization procedures even if only von Neumann (orthogonal) measurements are used.
Using more general measurements (POVM) the minimization problem gets increasingly demanding~\cite{galveEPL},
 with the consequence that there are only few general
results. Discord analytical expressions have been obtained only for certain
classes of two-qubit states, such as Bell-diagonal \cite{Luo2008PRA},
rank-2 \cite{Cen,galveEPL}  and X  \cite{Ali2010PRA} states and, for the case of
continuous variable, for Gaussian states \cite{paris-adesso}. To overcome
this drawback, geometric quantum discord $D_\mathrm{g}$ has been introduced
based on the square norm (Hilbert-Schmidt) distance between the system state and
its closest classical state and it has been used to evaluate quantum
correlations present in an arbitrary two-qubit state \cite{dakic2010PRL}.
Quantitative comparisons between REB and geometric discord have been reported
\cite{batle,adesso,sen} and their dynamics have been also compared, revealing
qualitative differences in their time behaviors \cite{Bellomo2011}. Geometric
measures of total correlations, $T_\mathrm{g}$, and classical correlations,
$C_\mathrm{g}$, have been also defined using the square norm distance with their
explicit expressions given only for Bell-diagonal states \cite{Bellomo2011}.

The aim of this paper is to discuss the role and use of the square norm distance to
quantify in a unified view various kinds of correlations in a two-qubit state.
To this purpose we will consider a quite general class of bipartite states for
which we will find explicit expressions for geometric quantifiers of total,
quantum and classical correlations. In analogy to what has been done with REB
correlation quantifiers, we will investigate the possibility to have closed
additive relations among correlation quantifiers based on the square norm.

The main point of this paper is to show that relevant qualitative differences are found
when one attempts to construct a unified view of correlations using different ways to measure
the distance between the relevant states. In the case of a quite general class of two-qubit states (X-states), we will be able to
analytically prove that, differently from what happens with REB distance, using the square norm distance measure total correlations cannot be in general separated in a quantum and a classical part satisfying an additive relation that involves only distances between states.

The paper is organized as follows, in Section II we introduce the framework of unified view of
correlations both for REB and its geometric counterpart, in Section III we present the class of states
on which we will base our study, X states, in Sections IV and V we obtain the pertinent
closest states. Finally, in Section VI we show under which conditions the closure of correlations is not satisfied and
study how often, and to what extent, this happens for X states.

\section{Correlations in a quantum state}
A natural and powerful way to quantify a given property of a quantum state
consists in exploiting the distance between the state itself and its closest
state without that property. Therefore, in this approach, it is necessary to
choose a suitable distance measure. In this section, we first briefly review the
correlation quantifiers defined by using relative entropy as a measure of
distance between states and we secondly describe the geometric correlation
quantifiers based on the square norm distance measure.

\subsection{Correlation quantifiers based on relative entropy}
\begin{figure}[t]
\includegraphics[width=0.25\textwidth]{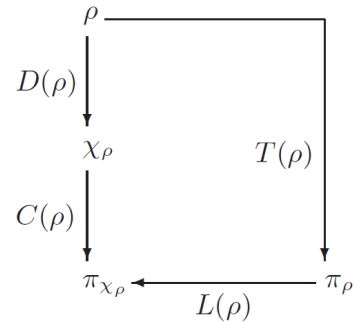}
\caption{\footnotesize Picture of the REB correlation quantifiers and the
relevant states (see Ref.~\cite{Modi2010PRL})}
\label{as}
\end{figure}

Given two arbitrary multipartite states $\rho,\sigma$, their relative entropy is
defined as $S(\rho\|\sigma)=-\mathrm{Tr}(\rho\log\sigma)-S(\rho)$, where
$S(\rho)=-\mathrm{Tr}(\rho\log\rho)$ is the von Neumann entropy. Using
relative entropy to quantify distances, the total correlations $T$ of a state
$\rho$ are defined by the distance between $\rho$ and the closest product
state $\pi_{\rho}=\sigma_A\otimes\sigma_B$ (with $\sigma_{A(B)}$ density matrices
for the subsystems), $T=S(\rho\|\pi_{\rho})$; discord (quantum
correlations) $D$ is the distance between $\rho$ and the closest classical
state $\chi_\rho=\sum_{i,j}p_{i,j}\ket{i}\bra{i}\otimes\ket{j}\bra{j}$ (with $\{\ket{i},\ket{j}\}$
independent local bases, and $p_{i,j}$ probabilities), $D=S(\rho\|\chi_{\rho})$, while classical correlations are
the distance between $\chi_\rho$ and its closest product state
$\pi_{\chi_{\rho}}$, $C=S(\chi_{\rho}\|\pi_{\chi_{\rho}})$ \cite{Modi2010PRL}.
We refer to these correlation quantifiers as relative entropy-based (REB)
quantifiers. In general, $T-(D+C)\neq 0$ but this difference is always equal to
another quantity $L$ defined in terms of relative entropy, that is
\begin{equation}\label{additive relation RE}
 T-D-C=-L,
\end{equation}
where $L=S(\pi_{\rho}\|\pi_{\chi_{\rho}})$. The validity of
Eq.~(\ref{additive relation RE}) is proved by the fact that all the involved REB
correlation quantifiers can be written as differences between von Neumann
entropies as \cite{Modi2010PRL}
\begin{eqnarray}\label{quantifiers}
T(\rho)&\equiv& S(\rho\|\pi_\rho)=S(\pi_\rho)-S(\rho),\nonumber\\
D(\rho)&\equiv& S(\rho\|\chi_\rho)=S(\chi_\rho)-S(\rho),\nonumber\\
C(\rho)&\equiv&
S(\chi_\rho\|\pi_{\chi_\rho})=S(\pi_{\chi_\rho})-S(\chi_\rho),\nonumber\\
L(\rho)&\equiv& S(\pi_{\rho}\|\pi_{\chi_{\rho}})=S(\pi_{\chi_\rho})-S(\pi_\rho).
\end{eqnarray}
These relations allow one to draw a simple diagram, proposed in
Ref.~\cite{Modi2010PRL} and displayed in Fig.~\ref{as}, where each line
refers to a kind of correlations, and where the direction of each arrow is
linked to the asymmetric definition of the relative entropy.
For bipartite systems, the quantity $L$ exactly quantifies the difference
between the REB discord $D$ of Eq.~(\ref{quantifiers}) and the original
definition $\delta$ \cite{zurek,henderson}: $\delta=D-L$. A closed
additivity relation eq.~(1) thus always holds among the REB correlation
quantifiers.

\subsection{Correlation quantifiers based on square norm}
As already noted, the REB correlation quantifiers have the drawback that
their analytical expressions are known only for certain classes of
states~\cite{Luo2008PRA,Cen,Ali2010PRA,paris-adesso} and require in general
numerical minimizations. A more manageable quantifier, named geometric quantum
discord, has been recently introduced for quantum correlations as the
square norm distance between the system state $\rho$ and its closest classical
state $\chi_\rho$ \cite{dakic2010PRL}
\begin{equation}\label{geometricdiscord}
    D_\mathrm{g}(\rho)=\|\rho-\chi_\rho\|^2,
\end{equation}
where $\|\cdot\|^2=\mathrm{Tr}(\cdot)^2$ is the square norm
distance in the Hilbert-Schmidt space
and $\chi_\rho$ has the form
$\chi_\rho=\sum_{i,j}p_{i,j}\ket{i}\bra{i}\otimes \rho^B_j$ (with $\{\ket{i}\}$
a local basis on part $A$, $\rho^B_j$ arbitrary states of part $B$ and $p_{i,j}$ probabilities). The advantage of this definition is that $D_\mathrm{g}(\rho)$ can be analytically evaluated for an arbitrary
two-qubit state and for some multipartite or higher dimensional systems more
easily than REB quantum discord \cite{dakic2010PRL,Luo2010PRA}. It is worth
noticing that the geometric definition of quantum discord of
Eq.~(\ref{geometricdiscord}) is equal to the one obtained in analogy to the
original definition of quantum discord and using the square norm distance
measure, that is
$D_\mathrm{g}(\rho)=\mathrm{min}_{\Pi^A}\|\rho-\Pi^A(\rho)\|^2$,
where $\Pi^A(\rho)$ is the classical state resulting after a von Neumann
measurement on part $A$ \cite{Luo2010PRA}.

Because geometric discord is useful to quantify quantum correlations in a
system, it looks a natural extension to use square norm also to define
quantifiers of total and classical correlations as \cite{Bellomo2011}
\begin{equation}\label{geometric total and classical}
    T_\mathrm{g}(\rho)\equiv\|\rho-\pi_\rho\|^2,\quad
C_\mathrm{g}(\rho)\equiv\|\chi_\rho-\pi_{\chi_\rho}\|^2,
\end{equation}
where $\pi_\rho$ and $\pi_{\chi_\rho}$ are, respectively, the product states
closest to $\rho$ and $\chi_\rho$ within the square norm distance measure.
We refer to the quantifiers based on square norm distance as geometric
correlation quantifiers. One can further define the quantity
\begin{equation}\label{L geometric}
L_\mathrm{g}(\rho)\equiv\|\pi_\rho-\pi_{\chi_\rho}\|^2,
\end{equation}
as the analogous of the REB quantity $L$ of Eq.~(\ref{quantifiers}).

In analogy of what happens for REB discord,
$D_\mathrm{g}$ can be written as a difference of purities $D_\mathrm{g}={\rm
Tr}(\rho-\chi_\rho)^2={\rm Tr}\rho^2-{\rm Tr}\chi^2_\rho$. In fact, analyzing the results
of Ref.~\cite{dakic2010PRL}, one can show that ${\rm Tr}(\rho\chi_{\rho})={\rm
Tr}(\chi_{\rho}^2)$. Differently, as we shall see in
Sec.~\ref{GeometricCorrelationQuantifiers}, the other geometric correlation
quantifiers $T_\mathrm{g}$ and $C_\mathrm{g}$ do not hold this property.

We shall now study the relationship among geometric correlation quantifiers
by firstly finding their explicit expressions for a quite general class of
bipartite states. We will also investigate if in general an additivity relation
analogous to that of Eq.~(\ref{additive relation RE}) is satisfied when
geometric correlation quantifiers are used.

\section{Two-qubit X states}
In this section we describe the class of two-qubit states we are going to use in our analysis. To
our aim it is useful to represent the states in the Bloch representation that,
for an arbitrary two-qubit state, is
\begin{equation}\label{bloch}
 \rho=\frac{1}{4}[\openone\otimes\openone+\sum_i x_i\sigma_i\otimes
\openone+\sum_i
y_i\openone\otimes\sigma_{i}+\sum_{i,j}T_{ij}\sigma_{i}\otimes\sigma_{j}],
\end{equation}
where $\openone$ is the $2\times2$ identity matrix, $\sigma_{i,j}$
($i,j=1,2,3$) are the three Pauli matrices,
$x_i=\mathrm{Tr}[\rho(\sigma_i\otimes\openone)]$ and
$y_i=\mathrm{Tr}[\rho(\openone\otimes\sigma_i )]$ are components of the local
Bloch vectors $\vec x=\{x_1,x_2,x_3\}$ and $\vec y=\{y_1,y_2,y_3\}$, and
$T_{ij}=\mathrm{Tr}[\rho(\sigma_i\otimes\sigma_j)]$ are components of the
correlation tensor $T$.

In particular, we put our attention to the class of X states, which are
those states having non-zero elements only along the main diagonal and
anti-diagonal of the density matrix. The general structure of an X density
matrix is thus, in the standard basis $\mathcal{B}=\{\ket{1}\equiv\ket{11},
\ket{2}\equiv\ket{10}, \ket{3}\equiv\ket{01}, \ket{4}\equiv\ket{00} \}$,
\begin{equation}\label{Xstates}
   \rho_X = \left(
\begin{array}{cccc}
  \rho_{11} & 0 & 0 & \rho_{14}e^{i \gamma_{14}}  \\
  0 & \rho_{22} & \rho_{23}e^{i \gamma_{23}} & 0 \\
  0 & \rho_{23} e^{- i \gamma_{23}} & \rho_{33} & 0 \\
  \rho_{14} e^{- i \gamma_{14}} & 0 & 0 & \rho_{44} \\
\end{array}
\right),
\end{equation}
where $\rho_{ij}$ ($i,j=1,2,3,4$) and $\gamma_{i,j}$ are all real, positive numbers. Bell
states, Werner states and Bell diagonal states belong to this class of states
\cite{BellomoASL}.  X-structure density matrices may
arise in a wide variety of physical situations and are also experimentally achievable \cite{Pratt2004PRL}. For example, X states are encountered as eigenstates in all the systems with odd-even symmetry like in the Ising and the XY models \cite{Fazio2002Nature}. Moreover, in many physical evolutions of open quantum systems an initial X structure is maintained in time \cite{Yu2007}.
The parameters of the Bloch representation of
Eq.~(\ref{bloch}) for an X state are thus expressed in terms of the the
density matrix elements of Eq.~(\ref{Xstates}) as \cite{RAU2009}
\begin{eqnarray}\label{X state Bloch}
&&x_3=\rho_{11}+\rho_{22}-\rho_{33}-\rho_{44}\,,\nonumber \\
&&y_3=\rho_{11}-\rho_{22}+\rho_{33}-\rho_{44}\,,\nonumber \\
&&T_{11}=2 \cos(\gamma_{14}) \rho_{14}+2 \cos(\gamma_{23}) \rho_{23}\,,\nonumber
\\
&&T_{12}=-2 \sin(\gamma_{14}) \rho_{14}+2 \sin(\gamma_{23})
\rho_{23}\,,\nonumber \\
&&T_{21}=-2 \sin(\gamma_{14}) \rho_{14}-2\sin(\gamma_{23}) \rho_{23}\,,\nonumber
\\
&&T_{22}=-2 \cos(\gamma_{14}) \rho_{14}+2 \cos(\gamma_{23})
\rho_{23}\,,\nonumber \\
&&T_{33}=\rho_{11}-\rho_{22}-\rho_{33}+\rho_{44},
\end{eqnarray}
with $x_1=x_2=y_1=y_2=T_{13}=T_{23}=T_{31}=T_{32}=0$.

We are interested in the explicit expressions of the geometric correlation
quantifiers for an X state. In order to obtain them, we first need to find the
relevant closest states when the distance is measured by the square norm.

\section{Closest product state}\label{par:Closest product state}
In this section we are interested in finding the product state closest to a
two-qubit X state in the square norm distance measure, with the aim to obtain
the geometric quantifier of total correlations.
Indicating with $\rho_A=\frac{1}{2}[\openone+\sum_i a_i\sigma_i]$ and
$\rho_B=\frac{1}{2}[\openone+\sum_i b_i\sigma_i]$ generic single-qubit states
with Bloch vectors, respectively, $\vec a=\{a_1,a_2,a_3\}$ and $\vec
b=\{b_1,b_2,b_3\}$, an arbitrary product state $\pi$ is given by their tensor
product as
\begin{eqnarray}\label{arbitraryproductstate}
\pi=\rho_A\otimes\rho_B&=&\frac{1}{4}\left[\openone\otimes\openone+\sum_i
a_i\sigma_i\otimes \openone+\sum_i
b_i\openone\otimes\sigma_{j}\right.\nonumber\\
&&\left.+\sum_{i,j}a_i b_j\sigma_{i}\otimes\sigma_{j}\right].
\end{eqnarray}
The distance $F$ between an arbitrary two-qubit state $\rho$ as given in
Eq.~(\ref{bloch}) and the product state $\pi$ using the square norm is then
\begin{eqnarray}\label{trace distance}
F=\mathrm{Tr}(\rho-\pi)^2&=&\frac{1}{4}\left[\sum_i (x_i-a_i)^2+\sum_i
(y_i-b_i)^2 \right.\nonumber\\
&&\left.+\sum_{i,j} (T_{ij}-a_ib_j)^2\right].
\end{eqnarray}
The explicit form of the product state $\pi_\rho$ closest to $\rho$ is
determined by the values of the variables $a_i,b_i$, as functions of the known
state parameters $x_i,y_i,T_{ij}$, giving the absolute minimum of the distance
$F$. Deriving $F$ with respect to $a_i$ and $b_j$ we construct the system
($i,j=1,2,3$)
\begin{eqnarray}\label{derivative system}
 a_i = \frac{x_i+\sum_j T_{ij}b_j}{1+\sum_j b_j^2} \,,\quad
  b_j = \frac{y_j+\sum_i T_{ij}a_i}{1+\sum_i a_i^2} .
\end{eqnarray}

In the case of X states, defined in Eqs.~(\ref{Xstates}) and (\ref{X state
Bloch}), the above system reduces to
\begin{eqnarray}\label{eqs12}
 a_1 &=& \frac{T_{11} b_1+T_{12} b_2}{1+\sum_j b_j^2} \,,\quad
 b_1 = \frac{T_{11} a_1+T_{21} a_2}{1+\sum_j a_j^2} \,,\nonumber \\
 a_2 &=& \frac{T_{21} b_1+T_{22} b_2}{1+\sum_j b_j^2}\,,\quad
 b_2 = \frac{T_{12} a_1+T_{22} a_2}{1+\sum_j a_j^2} \,,\nonumber\\
  a_3 &=& \frac{x_3+ T_{33}b_3}{1+\sum_j b_j^2} \,,\quad
  b_3 = \frac{y_3+T_{33}a_3}{1+\sum_i a_i^2} .
\end{eqnarray}
It can be shown (see appendix A) that the absolute minimum of $F$ is obtained by
putting $a_1=a_2=b_1=b_2=0$ and by taking the solutions for $a_3,b_3$ of
the system
\begin{eqnarray}\label{a3b3}
 a_3 = \frac{x_3+ T_{33}b_3}{1+ b_3^2} \,,\quad
  b_3 = \frac{y_3+T_{33}a_3}{1+ a_3^2} .
\end{eqnarray}
Indicating with $\bar{a}_3$ and $\bar{b}_3$ the solutions of
Eqs.~(\ref{a3b3}) and substituting in Eq.~(\ref{arbitraryproductstate}), the
product state closest to an X state has the form
\begin{equation}\label{product state to x state}
   \pi_{\rho_X}=\frac{1}{4}[\openone\otimes\openone+\bar{a}_3\sigma_3\otimes
\openone+ \bar{b}_3\openone\otimes\sigma_{3}+\bar{a}_3
\bar{b}_3\sigma_{3}\otimes\sigma_{3}].
\end{equation}
An important point, as we shall see later in the paper, is that the
parameters $\bar{a}_3$, $\bar{b}_3$ only depend on $x_3,y_3,T_{33}$, while
$T_{11},T_{12},T_{21},T_{22}$ do not play any role.

We observe that, in general, $\bar{a}_3\neq x_3$ and $\bar{b}_3\neq y_3$, that
means that the product state closest to an X state in the square norm distance
measure is not given by the product of its marginals
$\mathrm{Tr}_B(\rho_X)\otimes\mathrm{Tr}_A(\rho_X)$, differently from what
happens for any quantum state when the distance is measured by the
relative entropy \cite{Modi2010PRL}.

${}$\newline

\section{Closest classical state and its closest product
state}\label{ClosestClassicalState}
We now face the problem of finding explicit expressions of the classical
state closest to a an X state in the square norm distance measure. To this aim,
we follow a reported procedure that permits to obtain the closest classical
state given an arbitrary two-qubit state \cite{dakic2010PRL}. We point out
that, although explicit expressions for geometric quantum discord for X states
have been already reported in the literature \cite{altintas2010}, this is not
the case for the expressions of the corresponding closest classical states. Here
we also give the product state closest to the obtained closest classical state,
required for calculating the geometric quantifier of classical correlations (see
Eq.~(\ref{geometric total and classical})).

The general result for geometric quantum discord of two-qubit states is
$D_\mathrm{g}(\rho)=\frac{1}{4}[\|\vec{x}\|^2+\|T\|^2-k_{max}]$, where $k_{max}$
is the largest eigenvalue of matrix $K=\vec{x}\vec{x}^T+TT^T$ ($T^T$ is the
transpose of matrix $T$) \cite{dakic2010PRL}. The eigenvalues of matrix $K$
for an X state, in terms of the density matrix elements, are
\begin{eqnarray}\label{eigenvalues of K}
    k_1&=&4(\rho_{14}+\rho_{23})^2\,,\quad
k_2=4(\rho_{14}-\rho_{23})^2\,,\nonumber \\
    k_3&=&2[(\rho_{11}-\rho_{33})^2+(\rho_{22}-\rho_{44})^2]\,.
\end{eqnarray}
We observe that $k_1$ is always larger than $k_2$, so that only two
distinct cases have to be separately treated, that is $k_1\leq k_3$ and
$k_1>k_3$.

The closest classical state $\chi_\rho$ is obtainable by a minimization procedure with respect to the
parameters $\vec{x},\vec{y},T$ of the original state $\rho$ expressed in
the Bloch representation \cite{dakic2010PRL}. In the following we use this
procedure to obtain, for these two cases, the explicit expressions of the
closest classical state and of its closest product state for an X state defined
by the parameters of the Bloch representation given in Eq.~(\ref{X state
Bloch}).

${}$\newline

\subsection{Case 1: $k_1\leq k_3$}
When the X state has density matrix elements such that the condition
$k_1\leq k_3$ is fulfilled, we find that its closest classical state in
the square norm distance measure has the form
\begin{equation}\label{classical state first case}
   \chi_{\rho_X}^{(1)}=\frac{1}{4}[\openone\otimes\openone+x_3\sigma_3\otimes
\openone+ y_3\openone\otimes\sigma_{3}+T_{33}\sigma_{3}\otimes\sigma_{3}],
\end{equation}
where the superscript $(1)$ refers to the case 1 ($k_1\leq k_3$) of our
analysis.

Seeing that $\chi_{\rho_X}^{(1)}$ is still an X state, we can apply the results
of Sec.~\ref{par:Closest product state} to calculate its closest product state
$\pi_{\chi_{\rho_X}}^{(1)}$. The diagonal elements of $\rho_X$ and
$\chi_{\rho_X}$ are equal, $(\rho_X^{(1)})_{ii}=(\chi_{\rho_X^{(1)}})_{ii}$,
and, as said before, the solutions $\bar{a}_3$ and $\bar{b}_3$ of
Eq.~(\ref{a3b3}) only depend on the components $x_3,y_3,T_{33}$ containing the
diagonal density matrix elements. As a consequence, for X states lying in this
case 1, the product state closest to $\chi_{\rho_X}^{(1)}$ coincides with the
product state closest to $\rho_X$ given in Eq.~(\ref{product state to x state}):
$\pi_{\chi_{\rho_X}}^{(1)}=\pi_{\rho_X}$.

\subsection{Case 2: $k_1> k_3$}
If the X state has density matrix elements such that $k_1> k_3$, we obtain
for the closest classical state
\begin{eqnarray}\label{classical state second case}
\chi_{\rho_X}^{(2)}&=&\frac{1}{4}\Big\{
\openone\otimes\openone+y_3\openone\otimes\sigma_{3}
   +\tilde{T}_{11}\sigma_1\otimes\sigma_1 \nonumber
\\&&+\tilde{T}_{12}\sigma_1\otimes\sigma_2+\tilde{T}_{21}
\sigma_2\otimes\sigma_1+\tilde{T}_{22}\sigma_2\otimes\sigma_2\Big\},
\end{eqnarray}
where the superscript $^{(2)}$ refers to the case 2 ($k_1> k_3$) of our analysis
and
\begin{eqnarray}
  \tilde{T}_{11}&=&
\frac{1}{2}\left\{T_{11}[1+\cos(\gamma_{14}+\gamma_{23})])-T_{21}\sin(\gamma_{14
}+\gamma_{23})\right\}
  \nonumber \\& =& \left[\cos (\gamma_{23})+\cos(\gamma_{14})\right]
\left(\rho_{14}+\rho_{23}\right)\, ,  \nonumber \\
  \tilde{T}_{12}&=&
\frac{1}{2}\left\{T_{12}[1+\cos(\gamma_{14}+\gamma_{23})])-T_{22}\sin(\gamma_{14
}+\gamma_{23})\right\}
  \nonumber \\& =& \left[\sin(\gamma_{23})-\sin(\gamma_{14})\right]
\left(\rho_{14}+\rho_{23}\right)\, , \nonumber\\
  \tilde{T}_{21}&=&
\frac{1}{2}\left\{T_{21}[1-\cos(\gamma_{14}+\gamma_{23})])-T_{11}\sin(\gamma_{14
}+\gamma_{23})\right\}
  \nonumber \\& =& -\left[\sin(\gamma_{23})+\sin(\gamma_{14})\right]
\left(\rho_{14}+\rho_{23}\right)\, ,\nonumber\\
 \tilde{T}_{22}&=&
\frac{1}{2}\left\{T_{22}[1-\cos(\gamma_{14}+\gamma_{23})])-T_{12}\sin(\gamma_{14
}+\gamma_{23})\right\}
  \nonumber \\& =&  \left[\cos(\gamma_{23})-\cos(\gamma_{14})\right]
\left(\rho_{14}+\rho_{23}\right).
\end{eqnarray}

Here again $\chi_{\rho_X}^{(2)}$ has an X structure, so that considerations made
in Sec.~\ref{par:Closest product state} apply when one looks for its closest
product state. In particular, solving Eq.~(\ref{a3b3}) for
$\chi_{\rho_X}^{(2)}$, the closest product state is found to be
\begin{equation}
 \pi_{\chi_{\rho_X}}^{(2)}=\frac{1}{4}[
\openone\otimes\openone+y_3\openone\otimes\sigma_{3}].
\end{equation}
For an X state belonging to this case 2 it results, differently from the
case 1 above, that the closest product state to $\rho_X$ is different from the
product state closest to $\chi_{\rho_X}^{(2)}$:
$\pi_{\chi_{\rho_X}^{(2)}}\neq\pi_{\rho_X}$.

\section{Geometric correlation quantifiers and their relations\label{GeometricCorrelationQuantifiers}}
After obtaining the relevant closest states to a X state, we are now able
to give the explicit expressions of the geometric quantifiers of the various
kinds correlations $T_\mathrm{g}$, $D_\mathrm{g}$ and $C_\mathrm{g}$ and to
investigate their relations.

From the findings of Sec.~\ref{par:Closest product state} on the closest product
state, it follows that the geometric quantifier of total
correlations defined in Eq.~(\ref{geometric total and classical}) for an X
state is
\begin{eqnarray}\label{X total correlations}
T_\mathrm{g}(\rho_X)&=&\frac{\left(x_3-\bar{a}_3\right)^2+\left(y_3-\bar{b}
_3\right)^2+
    \left(T_{33}-\bar{a}_3\bar{b}_3\right)^2}{4}\nonumber \\
    &&+\frac{1}{4}(T_{11}^2+T_{12}^2+T_{21}^2+T_{22}^2),
\end{eqnarray}
for both cases 1 and 2.
We point out that in general ${\rm Tr}(\rho\pi_{\rho})\neq{\rm
Tr}(\pi^2_{\rho})$ and therefore $T_\mathrm{g}(\rho_X)\neq{\rm
Tr}(\rho_X^2)-{\rm Tr}(\pi^2_{\rho_{X}})$. Thus, total correlations
measured by the square norm distance are not expressible as a
difference between purities and they are not suitable to be represented by an
arrow from a state to another, as instead happens for REB correlation
quantifiers.

Concerning the geometric quantifiers of quantum and classical correlations,
respectively $D_\mathrm{g}$ and $C_\mathrm{g}$, we can obtain their explicit
expressions by the results of Sec.~\ref{ClosestClassicalState} on closest
classical state and its closest product state, distinguishing the two cases
$k_1\leq k_3$ (case 1) and $k_1> k_3$ (case 2).

For X states belonging to the case 1 ($k_1\leq k_3$), the geometric
quantifiers of quantum and classical correlations defined in
Eqs.~(\ref{geometricdiscord}) and (\ref{geometric total and classical}) are
\begin{eqnarray}\label{geometric quantifiers first case}
 D_\mathrm{g}^{(1)}&=&\frac{1}{4}(T_{11}^2+T_{12}^2+T_{21}^2+T_{22}^2)
=2(\rho_{14}^2+\rho_{23}^2),\nonumber \\
C_\mathrm{g}^{(1)}&=&\frac{\left(x_3-\bar{a}_3\right)^2+\left(y_3-\bar{b}
_3\right)^2+ \left(T_{33}-\bar{a}_3\bar{b}_3\right)^2}{4}.
\end{eqnarray}
Analogously to $T_\mathrm{g}$,
$C_\mathrm{g}=\|\chi_\rho-\pi_{\chi_\rho}\|^2$ is not expressible as a
difference between the purities of the states $\chi_\rho$ and $\pi_{\chi_\rho}$
and it is not representable in the same spirit of Eq.~(\ref{quantifiers}) by an
arrow. Regarding the quantity $L_\mathrm{g}$ defined in Eq.~(\ref{L geometric}),
we know from Sec.~\ref{ClosestClassicalState} that, in the case 1,
$\pi_{\chi_{\rho_X}}^{(1)}=\pi_{\rho_X}^{(1)}$ so that $L_g^{(1)}=0$. From
Eqs.~(\ref{X total correlations}) and (\ref{geometric quantifiers first case})
we then observe that, in this case, among the geometric quantifiers of
correlations the additivity relation
\begin{equation}
 T_\mathrm{g}=D_\mathrm{g}^{(1)}+C_\mathrm{g}^{(1)}
\end{equation}
holds, analogous to the one of Eq.~(\ref{additive relation RE}) for REB correlation quantifiers with $L=0$.

\begin{figure}[t]
\includegraphics[width=0.46\textwidth]{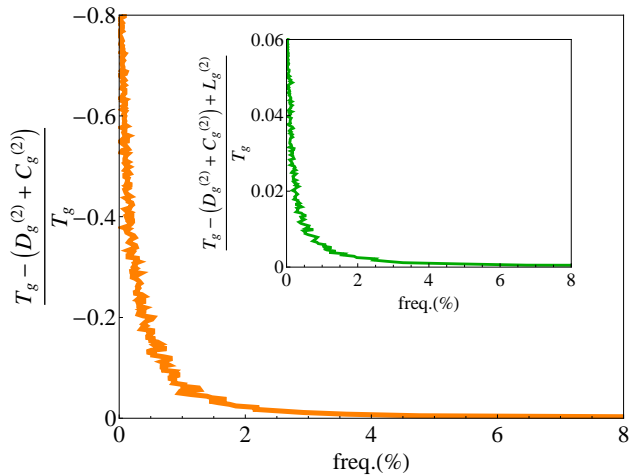}
\caption{\label{fig:class2}\footnotesize (Color online). Values of the
relative difference
$(T_\mathrm{g}-D_\mathrm{g}^{(2)}-C_\mathrm{g}^{(2)})/T_\mathrm{g}$ of the
geometric correlation quantifiers as a function of the probability of their
occurrence for a two-qubit X state belonging to the case 2 $k_1> k_3$. Inset:
Values of
$(T_\mathrm{g}+L_\mathrm{g}^{(2)}-D_\mathrm{g}^{(2)}-C_\mathrm{g}^{(2)}
)/T_\mathrm{g}$ as a function of the probability of their occurrence for a
two-qubit X state belonging to the same case 2. A number of $10^4$ random
density matrices has been produced.}
\end{figure}

On the other hand, for X states belonging to the case 2 ($k_1> k_3$) the
geometric quantifiers of quantum and classical correlations result to be
\begin{eqnarray}\label{geometric quantifiers second case}
    D_\mathrm{g}^{(2)}&=&(\rho_{14}-\rho_{23})^2
+\frac{1}{2}[(\rho_{11}-\rho_{33})^2 +
    (\rho_{22}-\rho_{44})^2]\,,\nonumber \\
   C_\mathrm{g}^{(2)}&=&(\rho_{14}+\rho_{23})^2,
\end{eqnarray}
where the expressions are given in terms of the density matrix elements
because they are simpler than those in terms of the Bloch parameters. From
Eqs.~(\ref{X total correlations}) and (\ref{geometric quantifiers second case})
we immediately notice that, in this case, an additivity relation analogous to
that of Eq.~(\ref{additive relation RE}) does not occur. In fact, we have
\begin{equation}\label{closcaso2}
 T_\mathrm{g}-D_\mathrm{g}^{(2)}-C_\mathrm{g}^{(2)}=
 \bar{a}_3^2\frac{\left(T_{33}-\bar{a}_3\bar{b}_3\right)^2-\left(1+\bar{b}
_3^2\right)}{4}.
\end{equation}
It can be shown (the proof is given in appendix B) that in general
$T_\mathrm{g}-D_\mathrm{g}^{(2)}-C_\mathrm{g}^{(2)}\le 0$ and the equality
holds if and only if $x_3+y_3 T_{33}=0$. In order to
estimate the typical value of $T_\mathrm{g}-D_\mathrm{g}^{(2)}-C_\mathrm{g}^{(2)}$, we have generated $10^4$
X-shaped random density matrices belonging to the case $k_1> k_3$, from
which we have then numerically obtained $\bar{a}_3$ and $\bar{b}_3$ as
solutions of Eq.~(\ref{a3b3}) and finally calculated all the correlations
Eqs.~(\ref{X total correlations}) and (\ref{geometric quantifiers second
case}). In Fig.~\ref{fig:class2}, we plot the values of
$(T_\mathrm{g}-D_\mathrm{g}^{(2)}-C_\mathrm{g}^{(2)})/T_\mathrm{g}$ versus the
(unnormalized) probability to have those amounts for an X state for which
$k_1>k_3$ (case 2). It is possible to see that there exist a nonnegligible amount of states for which
this difference is significantly different from zero.

In analogy to what happens for REB correlations in
Eq.(\ref{additive relation RE}), one may however wonder if the quantity
$L_\mathrm{g}$ of Eq.~(\ref{L geometric}) can be used to close the loop of
geometric correlations. We first observe that, as we know from
Sec.~\ref{ClosestClassicalState}, in this case 2
$\pi_{\rho_X}\neq\pi_{\chi_{\rho_X}}^{(2)}$ and then $L_\mathrm{g}^{(2)}\neq 0$.
In particular, it is
\begin{equation}
L_\mathrm{g}^{(2)}=\bar{a}_3^2\frac{\left(T_{33}-\bar{a}_3\bar{b}
_3\right)^2+\left(1+\bar{b}_3^2\right)}{4},
\end{equation}
so that
\begin{equation}
 T_\mathrm{g}-D_\mathrm{g}^{(2)}-C_\mathrm{g}^{(2)}+L_\mathrm{g}^{(2)}=
 \frac{\bar{a}_3^2}{2}\left(T_{33}-\bar{a}_3\bar{b}_3\right)^2.
\end{equation}
As a consequence, unless $\bar{a}_3=0$ which would also imply
$L_\mathrm{g}^{(2)}=0$, it is impossible to have a closed additive relation
among the different kinds of correlations when measured by the square norm. In
other words, geometric quantifiers of correlations can be cast in a closed
additivity relation only for X states with $L_\mathrm{g}$ equal to zero.
To quantitatively investigate this aspect, in the inset of
Fig.~\ref{fig:class2}, for the same $10^4$ randomly generated states as before,
we plot values of
$(T_\mathrm{g}+L_\mathrm{g}^{(2)}-D_\mathrm{g}^{(2)}-C_\mathrm{g}^{(2)}
)/T_\mathrm{g}$ as a function of the probability of their occurrence for a
two-qubit X state belonging to the same case 2. Even if the violation is in
general small, it is possible to find states for which it is meaningful. An important subclass of the X states is given by the Bell-diagonal states, for which a closed additivity relation holds among the correlation quantifiers, as it is discussed in appendix C.

We here point out that the numerical results above are useful to further
highlight the qualitative conceptual aspect on the impossibility to find in
general a closed relation among the geometric correlation quantifiers. For
example, there could be states, outside the class of X states, for which the
differences $T_\mathrm{g}-D_\mathrm{g}^{(2)}-C_\mathrm{g}^{(2)}$ and
$T_\mathrm{g}+L_\mathrm{g}^{(2)}-D_\mathrm{g}^{(2)}-C_\mathrm{g}^{(2)}$ are
larger than those we numerically find here for X states. Moreover, as discussed
before, another critical point arising when using square norm distance measure
to quantify correlations is the absence of a ``direction'' in the correlation
measures. While REB correlation quantifiers can be represented by arrows with
precise directions from a state to another one, as Eqs.~\ref{quantifiers} and
Fig.~\ref{as} clearly indicate, this is no longer true when geometric
quantifiers of correlations are taken into account.

\section{Physical model}
\begin{figure}
\includegraphics[width=0.46\textwidth]{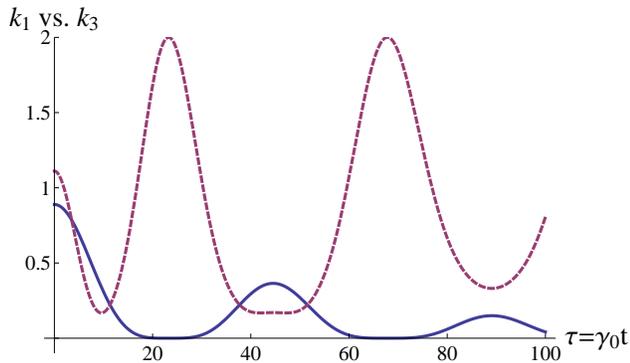}
\caption{\label{fig:3}\footnotesize (Color online). $k_1$ (blue solid line) versus $k_3$ (purple dashed line) as a function of the dimensionless time $\gamma_0t$ for $\lambda=0.01\gamma_0$ and starting from the state $\ket{\Psi}=\sqrt{1/3}\ket{00}+\sqrt{2/3}\ket{11}$.}
\end{figure}

In this section we consider a specific physical model where the two state space zones $k_1\leq k_3$ and $k_1> k_3$ above investigated can be dynamically connected. In particular, we take two noninteracting qubits, $A$ and $B$, embedded in separated cavities and subject to
a non-Markovian dynamics, as already reported in Ref.~\cite{Bellomo2007PRL}. Each qubit
interacts  only and independently with its local environment, so that the total Hamiltonian is $H_\mathrm{tot}=H_A+H_B$. The single ``qubit+reservoir'' Hamiltonian $H_S$ ($S=A,B$) is given by ($\hbar=1$)
\begin{equation}\label{Hamiltonian}
H_S=\omega_0^S \sigma_+^S\sigma_-^S+\sum_k\omega_k b_k^{S\,\dag } b_k^S+\sum_k(g_k^S\sigma_+^S b_k^S+g_k^{S\,\ast}\sigma_-^S b_k^{S\,\dag}),
\end{equation}
where $\omega_0^S$ is the transition
frequency of the two-level system (qubit) $S$, $\sigma_\pm^S$ are
the system raising and lowering operators, the index $k$
labels the field modes of the reservoir $S$ with frequencies
$\omega_k$, $b_k^{S\,\dag} $, $b_k^S $ are the modes creation and
annihilation operators and $g_k^S$ the coupling constants. We
will consider the case where both Hamiltonians $H_A$ and $H_B$ have the same parameters. The
Hamiltonian of Eq.~(\ref{Hamiltonian}) may
represent a qubit made by the excited and ground electronic
states of a two-level atom interacting with a reservoir given by
the quantized modes of a high-$Q$ cavity \cite{kuhr2007APL} and it can also be implemented by superconducting Josephson qubits in the framework of circuit QED \cite{wallraff2009PRL} and by entangled polarization photons in an all-optical setup \cite{almeida2007Science}. If the two-qubit state has initially an X structure, this is maintained during the dynamics locally governed by the Hamiltonian of Eq.~(\ref{Hamiltonian}) and the density matrix elements at time $t$, in the same basis as in Eq.~(\ref{Xstates}), are \cite{Bellomo2007PRL}
\begin{eqnarray}\label{rototdiag}
\rho_{11}(t)&=&\rho_{11}(0)P_t^2,\nonumber\\
\rho_{22}(t)&=&\rho_{22}(0)P_t+\rho_{11}(0)P_t(1-P_t),\nonumber\\
\rho_{33}(t)&=&\rho_{33}(0)P_t+\rho_{11}P_t(1-P_t),\nonumber\\
\rho_{44}(t)&=&1-[\rho_{11}(t)+\rho_{22}(t)+\rho_{33}(t)],\nonumber\\
|\rho_{14}(t)|&=&|\rho_{14}(0)|P_t,\quad |\rho_{23}(t)|=|\rho_{23}(0)|P_t,
\end{eqnarray}
where $P_t=\mathrm{e}^{-\lambda t}\left[ \cos \left(\frac{d
t}{2}\right)+\frac{\lambda}{d}\sin \left(\frac{d t}{2}\right)
\right]^2$, with $d=\sqrt{2\gamma_0 \lambda-\lambda^2}$. The parameter $\lambda$ represents the spectral width of the coupling  while $\gamma_0$ is the spontaneous emission rate of the qubit (atom). Under this evolution the two-qubit states may cross the two zones $k_1\leq k_3$ and $k_1> k_3$ of Sec.~\ref{ClosestClassicalState}. This can be explicitly seen by choosing, for example, the initial state $\ket{\Psi}=\sqrt{1/3}\ket{00}+\sqrt{2/3}\ket{11}$ and $\lambda=0.01\gamma_0$ (strong coupling regime). In Fig.~\ref{fig:3} we plot $k_1$ and $k_3$ as functions of the dimensionless time $\gamma_0t$. The plot clearly displays that there are time regions when $k_1$ is larger than $k_3$ and vice versa.

\section{Conclusions}
The distance between a quantum state and its closest states without
certain properties has been employed in the literature to quantify in a unified
view the various kinds of correlations present in an arbitrary multipartite
quantum system. Using relative entropy as distance measure, an additivity
relation among total $T$, quantum $D$ and classical $C$ correlations
holds~\cite{Modi2010PRL}. This additivity relation is of the kind $T=D+C-L$ and
contains a quantity $L$ defined as the relative entropy-based (REB) distance
between two particular closest product states linked to the system state
\cite{Modi2010PRL}.

In this paper we have investigated how a unified view of various kinds of
correlations works in terms of a different distance measure. Among the possible
suitable distance measures, we have considered the square norm (Hilbert-Schmidt)
distance, inspired by the fact that it has been already exploited to define
geometric quantum discord $D_\mathrm{g}$ to quantify quantum correlations
present in a state \cite{dakic2010PRL}. Using the square norm, we have then
considered the geometric quantifiers for classical $C_\mathrm{g}$ and total
$T_\mathrm{g}$ correlations, recently introduced \cite{Bellomo2011}, and defined
a quantity $L_\mathrm{g}$, analogous to the REB quantity $L$. We have given
explicit expressions of the geometric correlation quantifiers for the class of
two-qubit X states $\rho_X$, by firstly finding the relevant closest states
linked to $\rho_X$. We have then analyzed the relationship among the various
correlation quantifiers and we have shown that there exist a subclass of X
states for which it is not possible to find a closed additivity relation of the
kind $T_\mathrm{g}=D_\mathrm{g}+C_\mathrm{g}-L_\mathrm{g}$. The additivity
relation holding when relative entropy is used as distance measure is therefore
not preserved when one quantifies in a unified way the different kinds of
correlations present in a quantum state by using the square norm. Moreover
we have explored numerically the abundance and importance of the nonadditivity of
geometric correlations and found that there is a nonnegligible amount of X states with
a meaningful deviation from additivity. Therefore, both the analytical proof given for a subclass of X states and the numerically evaluated occurrence among X-states show that for a two-qubit random state, a closed relation among geometric correlation quantifiers needs not to be satisfied.

The results of this paper seem to confirm that different distance measures
cannot serve just as well to quantify the various kinds of correlations present
in a quantum state. In this sense, in Ref.~\cite{Bellomo2011} it had been
shown that, when one compares the dynamics of correlation quantifiers based on
relative entropy with that of correlation quantifiers based on square norm
distance, different qualitative behaviors may occur. These findings recall the
known result in the entanglement theory that different entanglement measures may
induce different orderings in the state space \cite{Eisert1999JMO}. The present
results give a clear indication that different distance measures (in particular
relative entropy and square norm) may not share important general qualitative
properties, like the occurrence of closed additivity relations among the various
correlation quantifiers.

\appendix

\section{}
In this appendix we show that the function $F$ of Eq.~(\ref{trace
distance}), measuring the distance between a two-qubit state $\rho$ and the
arbitrary product state $\rho_A\otimes\rho_B$ of
Eq.~(\ref{arbitraryproductstate}), has an absolute minimum for
$a_1=a_2=b_1=b_2=0$ and for $a_3=\bar{a}_3$, $b_3=\bar{b}_3$ solutions of
Eqs.~(\ref{a3b3}) when an X state is considered.

Let us consider the function $F$ of Eq.~(\ref{trace distance}), in the case of X
states. It can be divided into two parts. We write $F=F_1+F_2$, where
\begin{eqnarray}
 F_1&=&\frac{1}{4}[a_1^2+b_1^2+a_2^2+b_2^2\nonumber\\&&+(T_{11}-a_1
b_1)^2+(T_{12}-a_1 b_2)^2\nonumber\\&&+(T_{21}-a_2 b_1)^2+(T_{22}-a_2
b_1)^2\nonumber\\&&+a_1^2 b_3^2+a_2^2 b_3^2+a_3^2 b_1^2+a_3^2 b_2^2],
\end{eqnarray}
and
\begin{equation}\label{f2}
 F_2=\frac{1}{4}[(x_3-a_3)^2+(y_3-b_3)^3+(T_{33}-a_3b_3)^2].
\end{equation}
We observe that $F_2$ only depends on $a_3,b_3$ and also that, if
$a_1=b_1=a_2=b_2=0$ gives a minimum for $F_1$, this occurs irrespectively of the
value of $a_3$ and $b_3$. Therefore, the absolute minimum of $F$ is
obtained for the values of variables giving the absolute minimum of $F_1$ and
$F_2$ separately. The absolute minimum of the convex function $F_2$ is just
obtained in the values $a_3=\bar{a}_3$ and $b_3=\bar{b}_3$ that are solutions of
Eqs.~(\ref{a3b3}). In order to verify that $F_1$ actually has a minimum in
$a_1=b_1=a_2=b_2=0$, it is enough to consider the part of function $F_1$
non-including $a_3,b_3$ (the remaining part is equal to zero for
$a_1=b_1=a_2=b_2=0$). Thus we focus on the function $f=f(a_1,b_1,a_2,b_2)$ given
by
\begin{eqnarray}
f&=&a_1^2+b_1^2+a_2^2+b_2^2+(T_{11}-a_1 b_1)^2+(T_{12}-a_1
b_2)^2\nonumber\\&&+(T_{21}-a_2 b_1)^2+(T_{22}-a_2 b_1)^2.
\end{eqnarray}

In order to show that $f$ has an absolute minimum in $a_1=b_1=a_2=b_2=0$, we
consider the difference $\Delta f=f(a_1,b_1,a_2,b_2)-f(0,0,0,0)$ between
the value of $f$ in any possible point and its value in $\{0,0,0,0\}$, that
is
\begin{eqnarray}\label{deltaf}
\Delta f&=&a_1^2+b_1^2+a_2^2+b_2^2+a_1^2b_1^2+a_2^2b_2^2+a_2^2b_1^2+a_1^2b_2^2
\nonumber\\
 &&-(T_{11}a_1b_1+T_{12}a_1b_2+T_{21}a_2b_1+T_{22}a_2b_2).
\end{eqnarray}
The last term of Eq.~(\ref{deltaf}) admits as a minimum value
$-(|T_{11}a_1b_1|+|T_{12}a_1b_2|+|T_{21}a_2b_1|+|T_{22}a_2b_2|)$. Furthermore,
since $|T_{ij}|\le 1$, the lower bound of this expression is
$-(|a_1b_1|+|a_1b_2|+|a_2b_1|+|a_2b_2|)$, so that
\begin{eqnarray}\label{df}
 \Delta f&\ge&
a_1^2+b_1^2+a_2^2+b_2^2+a_1^2b_1^2+a_2^2b_2^2+a_2^2b_1^2+a_1^2b_2^2\nonumber\\
 &&-(|a_1b_1|+|a_1b_2|+|a_2b_1|+|a_2b_2|).
\end{eqnarray}
We now observe that
\begin{eqnarray}\label{forma1}
 a_1^2+b_1^2+a_2^2+b_2^2&=&(|a_1|-|b_1|)^2+(|a_2|-|b_2|)^2\nonumber\\&&+2
|a_1b_1|+2 |a_2b_2|
\end{eqnarray}
and also
\begin{eqnarray}\label{forma2}
 a_1^2+b_1^2+a_2^2+b_2^2&=&(|a_1|-|b_2|)^2+(|a_2|-|b_1|)^2\nonumber\\&&+2
|a_1b_2|+2 |a_2b_1|
\end{eqnarray}
Putting these expressions in Eq.~(\ref{df}), we obtain
\begin{eqnarray}
 \Delta f&\ge&
\frac{1}{2}[(|a_1|-|b_1|)^2+(|a_2|-|b_2|)^2+(|a_1|-|b_2|)^2\nonumber\\
 &&+(|a_2|-|b_1|)^2]+ a_1^2b_1^2+a_2^2b_2^2+a_2^2b_1^2+a_1^2b_2^2.
\end{eqnarray}
Therefore $\Delta f$ is always greater than zero
unless $a_1=b_1=a_2=b_2=0$. {\bf QED}.

\section{}
In this appendix we want to prove that the quantity
$h=\bar{a}_3^2\left[\left(T_{33}-\bar{a}_3\bar{b}_3\right)^2-\left(1+\bar{b}
_3^2\right)\right]$, where $\bar{a}_3,\bar{b}_3$ are solutions of
Eqs.~(\ref{a3b3}), that quantifies the difference
$T_\mathrm{g}-D_\mathrm{g}^{(2)}-C_\mathrm{g}^{(2)}=h/4$ given in
Eq.~(\ref{closcaso2}), satisfies the inequality $h\leq0$ with the equality
verified if and only if $x_3+y_3 T_{33}=0$. To this aim, being
$1+\bar{b}_3^2\ge 1$, it will be sufficient to prove that
$\left|T_{33}-\bar{a}_3\bar{b}_3\right|\le 1$ and successively to find the
condition for which the upper bound 1 is achieved.

Let us start noticing that $\bar{a}_3,\bar{b}_3$ give the minimum of the
function $F_2$ of Eq.~(\ref{f2}). Thus, $F_2(a_3,b_3)\ge F_2(\bar
a_3,\bar b_3)\ge(T_{33}-\bar{a}_3\bar{b}_3)^2$.
Then, if in  Eq.~(\ref{f2}) we replace $a_3,b_3$ with $x_3,y_3$, the inequality
\begin{equation}\label{ineq}
 (T_{33}-x_3 y_3)^2\ge (T_{33}-\bar{a}_3 \bar{b}_3)^2
\end{equation}
holds.
Using Eqs.~(\ref{X state Bloch}) of an X state, we have
\begin{equation}
 T_{33}-x_3 y_3=p_1-t_1^2-(p_2-t_2^2),
\end{equation}
where $p_1=\rho_{11}+\rho_{44}$, $p_2=1-p_1=\rho_{22}+\rho_{33}$,
$t_1=\rho_{11}-\rho_{44}$, and $t_2=\rho_{22}-\rho_{33}$.
$T_{33}-x_3 y_3$ reaches its maximum value for $t_1=0$ and $t_2=p_2$, while
its minimum value is found for $t_1=p_1$ and $t_2=0$. In any case
$(T_{33}-x_3 y_3)_{\max}=p_1+p_2^2-p_2\le 1-p_2$ and $(T_{33}-x_3
y_3)_{\min}=p_1-p_1^2-p_2\ge -1+p_1$. Then $\left|T_{33}-x_3 y_3\right|\le 1$
and, according to inequality~(\ref{ineq}),
$\left|T_{33}-\bar{a}_3\bar{b}_3\right|\le 1$, as we wanted to prove.

The cases such that $h=\bar{a}_3^2
[\left(T_{33}-\bar{a}_3\bar{b}_3\right)^2-\left(1+\bar{b}_3^2\right)]= 0$
correspond to i) $\bar{a}_3=0$ or to ii) $\bar{b}_3=0$ and $|T_{33}|=1$. From
Eqs.~(\ref{a3b3}), we can notice that case (i) implies $b_3=y_3$ and then
$x_3+y_3 T_{33}=0$; case (ii) instead implies that $\bar{a}_3=x_3=\pm y_3$ where
$\pm y_3$ correspond to $T_{33}=\mp 1$ respectively, so that again $x_3+y_3
T_{33}=0$.

\section{}
In this appendix we discuss the properties of a particular subclass of X states, namely the Bell-diagonal states, also called states with maximally mixed marginals \cite{Luo2008PRA}. Looking at the matrix form of an X state given in Eq.~(\ref{Xstates}), Bell-diagonal states have diagonal elements $\rho_{11}=\rho_{44}$, $\rho_{22}=\rho_{33}$ and non-diagonal density matrix elements real, that is $\gamma_{14},\gamma_{23}=0,\pi$. Therefore, the parameters of their Bloch representation are $x_3=y_3=T_{12}=T_{21}=0$, $T_{11}=2\mathrm{e}^{\mathrm{i}\gamma_{14}}[\rho_{14}+\mathrm{e}^{\mathrm{i}
(\gamma_{14}-\gamma_{23})}\rho_{23}]$, $T_{22}=2\mathrm{e}^{\mathrm{i}\gamma_{14}}[\mathrm{e}^{\mathrm{i}(\gamma_{14}
-\gamma_{23})}\rho_{23}-\rho_{14}]$ and $T_{33}=2(\rho_{11}-\rho_{22})$, giving
a Bloch representation \cite{Luo2008PRA}
\begin{equation}\label{Belldiagonalstate}
\rho^\mathrm{B}=[\openone\otimes\openone+\sum_{i=1}^3T_{ii}
\sigma_i\otimes\sigma_i]/4.
\end{equation}
Bell-diagonal states have the peculiar property that the quantity $L$ of Eq.~(\ref{quantifiers}) is zero ($L=0$), so that $T=D+C$ for REB correlation quantifiers \cite{Modi2010PRL}. Indeed, these states present the same closed additivity relation even for the geometric correlation quantifiers of Eqs.~(\ref{geometricdiscord}) and (\ref{geometric total and classical}), that is $T_\mathrm{g}=D_\mathrm{g}+C_\mathrm{g}$ \cite{Bellomo2011} with $L_\mathrm{g}=0$.

Bell-diagonal states $\rho^\mathrm{B}$ have $x_3=y_3=0$ and it is possible to show that the corresponding solutions of Eqs.~(\ref{a3b3}) are $\bar{a}_3=\bar{b}_3=0$ \cite{Bellomo2011}, so that the closest product state using the square norm, in this case, reduces to the product of the marginals $\pi_{\rho^B}=(\openone/2)\otimes(\openone/2)$.

Regarding the closest classical state and its closest product state (see Sec.~\ref{ClosestClassicalState}), for Bell-diagonal states it occurs that, if $\gamma_{14}=\gamma_{23}$ then $k_{ii}=T_{ii}^2$, while if $\gamma_{14}=\pi-\gamma_{23}$ one has $k_{11}=T_{22}^2$, $k_{22}=T_{11}^2$ and always $k_{33}=T_{33}^2$. The two cases $k_1\leq k_3$ and $k_1>k_3$, obtained from Eq.~(\ref{eigenvalues of K}), thus involve direct comparisons among absolute values of the components $T_{11},T_{22},T_{33}$ of the correlation tensor. Then, if a Bell-diagonal state belongs to the case 1 ($k_1\leq k_3$), being $x_3=y_3=0$, the closest classical state of Eq.~(\ref{classical state first case}) reduces to $\chi_{\rho^\mathrm{B}}^{(1)}=\frac{1}{4}[\openone\otimes\openone+T_{33}\sigma_{3}\otimes\sigma_{3}]$ with $T_{33}^2=k_3$. On the other hand, for a Bell-diagonal state belonging to the case 2 ($k_1>k_3$), being $y_3=0$ and $\gamma_{14},\gamma_{23}=0,\pi$, the closest classical state of Eq.~(\ref{classical state second case}) reduces either to
$\chi_{\rho^\mathrm{B}}^{(2)}=\frac{1}{4}[\openone\otimes\openone+T_{11}\sigma_{1}\otimes\sigma_{1}]$ if $\gamma_{14}=\gamma_{23}$ with $T_{11}^2=k_1$, or to $\chi_{\rho^\mathrm{B}}^{(2)}=\frac{1}{4}[\openone\otimes\openone+T_{22}\sigma_{2}\otimes\sigma_{2}]$ if $\gamma_{14}=\pi-\gamma_{23}$ with $T_{22}^2=k_1$. Since we have already seen that the closest product state to $\rho^\mathrm{B}$ is given by the product of its
marginals, the same happens for the product state closest to the closest classical state $\chi_{\rho^\mathrm{B}}^{(1)}$ or $\chi_{\rho^\mathrm{B}}^{(2)}$.

For a Bell-diagonal state the explicit expression of geometric quantum discord can be written as $D_\mathrm{g}(\rho^\mathrm{B})=\frac{1}{4}[T_{11}^2+T_{22}^2+T_{33}^2-T^2]$ \cite{dakic2010PRL}, where $T\equiv\mathrm{max}\{|T_{11}|,|T_{22}|,|T_{33}|\}$. The geometric quantifier of total correlations of Eq.~(\ref{X total correlations}) reduces to
$T_\mathrm{g}(\rho^\mathrm{B})=(T_{11}^2+T_{22}^2+T_{33}^2)/4$ with the geometric quantifier of classical correlations given by $C_\mathrm{g}(\rho^\mathrm{B})=T^2/4$ \cite{Bellomo2011}, so that it is always $T_\mathrm{g}=D_\mathrm{g}+C_\mathrm{g}$. In particular, for Bell-diagonal states belonging to the case 2, the second member of Eq.~(\ref{closcaso2}) is zero because $x_3+y_3 T_{33}=0$, being $x_3=y_3=0$.

\end{document}